\documentclass[osajnl,twocolumn,showpacs,superscriptaddress,10pt]{revtex4-1} 
\usepackage{amsmath,amssymb,graphicx}
\begin{document}

\title{Simultaneous measurement of mass and rotation of trapped absorbing
particles in air}

\author{Sudipta K. Bera\textsuperscript{\dag{}}}
\author{Avinash Kumar\textsuperscript{\dag{}} }
\author{Souvik Sil }
\author{Tushar Kanti Saha}
\author{Tanumoy Saha}
\author{Ayan Banerjee}

\affiliation{Department of Physcial Sciences, Indian Institute of Science Education and Research, Kolkata, India - 741246}
\email{Corresponding author: ayan@iiserkol.ac.in}
\collaboration{\dag These authors have equal contributions in the work.}

\begin{abstract}
We trap absorbing micro-particles in air by photophoretic forces generated
using a single loosely focused Gaussian trapping beam.
We measure a component of the radial Brownian motion of a trapped particle cluster and determine the power spectral density, mean squared displacement,
and normalized position and velocity autocorrelation functions in order to characterize the photophoretic body force in a quantitative fashion for the first time. The trapped particles also undergo spontaneous rotation due to the action of this force. This is evident from the spectral density that displays clear peaks at the  rotation and the particles' inertial resonance frequencies. We fit the spectral density to the well-known analytical function derived from the Langevin equation, measure the resonance and rotation frequencies and determine values for particle mass that we verify at different trapping laser powers with reasonable accuracy. 
\end{abstract}

\maketitle

Photophoretic forces \cite{jovanovic2009photophoresis}
have provided an alternate route for trapping absorbing mesoscopic
particles in air, as these forces, having a thermal origin, are  almost
four orders of magnitude higher than optical radiation pressure or
dipole forces \cite{Desyatnikov09}, when acting on particles of the
same size. Such forces can therefore balance gravity, and recently,
extensive use has been made of them to trap \cite{braun2013optically,jauffred2015optical},
controllably manipulate \cite{shvedov2010giant,Shvedov:09,zhang2012observation},
or even rotate \cite{lin2014optical} particles in air using rather simple
experimental configurations and without the use of tight focusing
objective lenses typically warranted in optical gradient force trapping. However, there has been very little attempt to quantify
the effects of these forces and observe their manifestations in the
Brownian motion of trapped particles in comparison to the extensively studied
problem of trapping using optical gradient forces.

\begin{figure}
\includegraphics[scale=0.25]{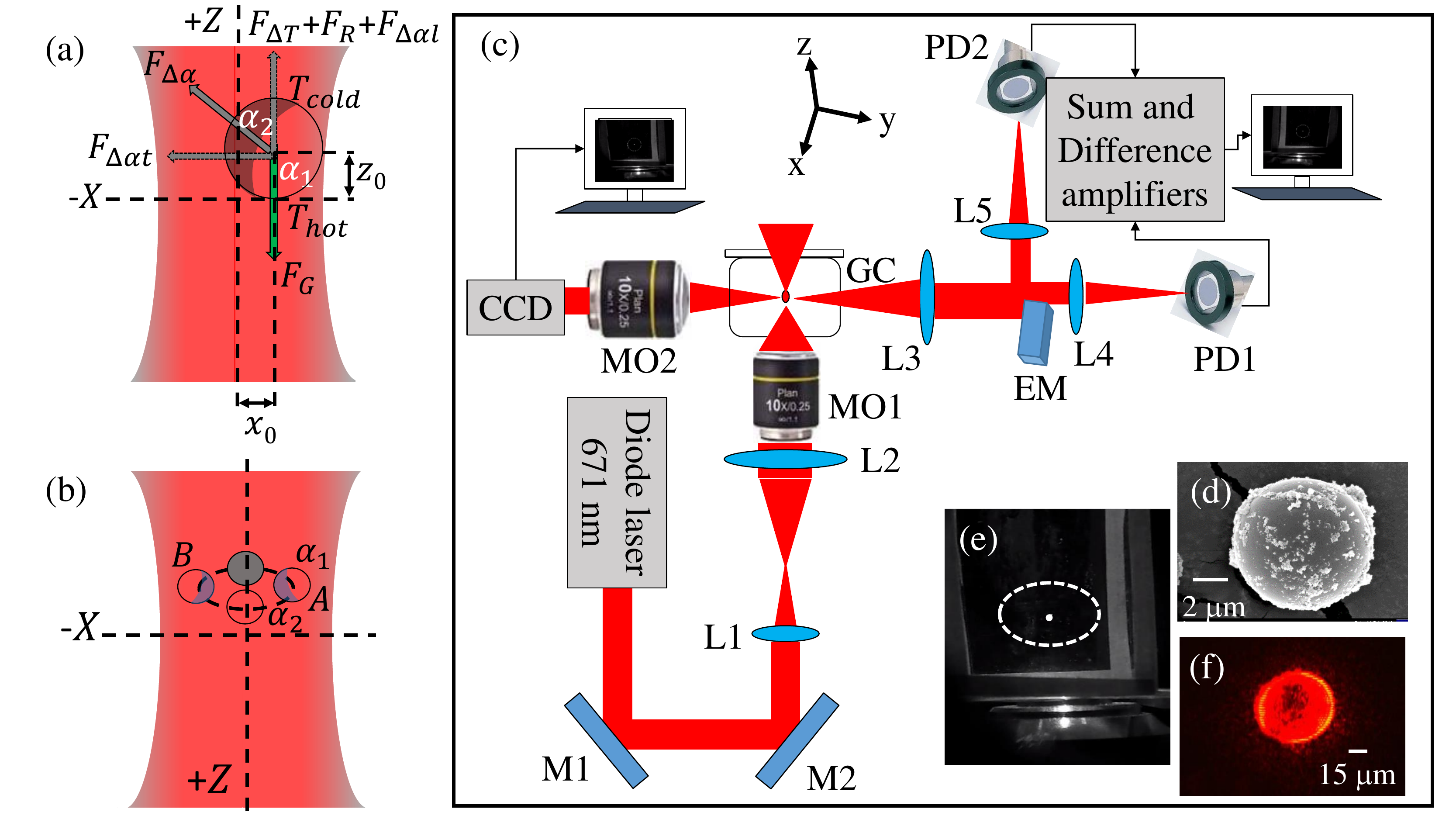}\caption{\label{fig1}(a). Schematic diagram of the forces acting on a trapped particle in our trapping configuration. The accommodation coefficients $\alpha_1$ and $\alpha_2$ are shown in (b) Demonstration of how the torque induced by $F_{\Delta\alpha T}$ on the particle causing it to rotate also results in a restoring force in the radial direction. The rotation causes the particle to flip which subsequently flips the direction of  $F_{\Delta\alpha T}$ between points A and B. (c) Schematic of the experiment. M1 and M2: plane mirrors, L1, L2, L3, L4 and L5: plano-convex lenses, MO1 and MO2: 10X objective lenses, EM: Edge mirror, PD1 and PD2: Photodiodes, GC: glass cuvette, C: Camera. (d) SEM image of a single coated silicon oxide sphere, the diameter being around 8 $\mu m$. (e) Image of a trapped cluster of spheres using scattering from the laser. (f) Zoomed in image of (e).}
\end{figure}

In this paper, we address this issue, and study the motion of Brownian
particles trapped under the influence of photophoretic forces. The
particles are absorbing in nature and trapped in a very simple experimental
set-up by a single Gaussian beam that is focused by a low magnification
microscope objective.  As shown in Fig.~ \ref{fig1}(a), we trap the particles in a vertical configuration,
i.e. with the particles falling under gravity ($-z$ direction) while
the laser beam travels in the $+z$ direction.
In this scenario, the particle is axially in equilibrium when gravity
is balanced by the action of the radiation pressure and photophoretic
forces, i.e.$F_{\Delta T}+F_{R}+F_{\Delta\alpha L}=F_{G}$ , where
$F_{G}$ is the force due to gravity, $F_{R}$ is the radiation pressure
force, $F_{\Delta T}$ is the photophoretic $\Delta T$
force arising due to difference of temperature on two opposite surfaces
of the particle ($T_{hot}-T_{cold}$), while $F_{\Delta\alpha L}$
is the longitudinal component of the photophoretic body force $F_{\Delta\alpha}$ that is generated due to the variation of the accommodation coefficient
$\alpha$ across the surface of the trapped particle ($\alpha_{1}$
and $\alpha_{2}$, with $\alpha_{1}>\alpha_{2}$). The direction of
this force is from $\alpha_{1}$to $\alpha_{2}$ \cite{jovanovic2009photophoresis}.
At atmospheric pressures, $F_{\Delta\alpha}$ dominates over $F_{\Delta T}$
\cite{Rohatschek1995}, and is therefore the dominant force balancing
gravity. Due to the action of these competing forces, the particle
is not necessarily trapped at the focus of the Gaussian beam, but
at an axial distance $z_{0}$ from the beam center where the net force
is zero \cite{zhang2012observation,lin2014optical}. The radial trapping
is solely achieved by the transverse component of the body force $F_{\Delta\alpha T}$
which is similar to optical gradient forces in our experimental configuration
since it is purely restoring in nature, there being no other balancing
forces. It is also understandable that since $F_{\Delta\alpha}$ is
a body-fixed force, its direction with respect to the gravity leads
to generation of a torque on the particle about the axis of $F_{G}$,
that leads to rotation of the trapped particle in the transverse direction
\cite{wurm2008}. In fact, it is this rotation that provides the restoring
force as the direction of $F_{\Delta\alpha}$ reverses from points
A to B in the rotation path shown in Fig.~ \ref{fig1}(b). Evidence
of such rotation was recently demonstrated in Ref. \cite{lin2014optical},
where the authors measured the rotation frequency in the time domain
using simple detectors.  However, other than rotation, the
trapped particle also undergoes Brownian motion, which may be difficult
to detect in the time domain due to the presence of the strong
intensity modulation of the scattered light due to rotation of the
particle. To resolve this issue, we use a position-sensitive detection system, and measure the Brownian motion of a trapped particle
cluster from the scattered light intensity. This allows us to
perform both time and frequency domain analyses to obtain interesting results which include estimates of the mass of the trapped particles. 

The experimental schematic is shown in Fig.~ \ref{fig1}(c). For the experiment, we coat commercial $SiO_{2}$ spheres (Sigma Aldrich, mean diameter between 9-13 $\mu m$, density 1100 $kg/m^{3}$) with $PbS$ by sonicating them in a sulfide salt solution followed by heat exposure in a furnace. A SEM image of the coated beads is shown in Fig.~ \ref{fig1}(d). The coating is not uniform, but in patches, and for determining mass of the particles, we consider an average thickness of around 100 nm which we determine by the total area covered by coated material compared to the area of the $SiO_{2}$ particles. The trapping laser is a diode laser at 671 nm with maximum power 300 mW which we couple into a 10X objective (MO1) using appropriate beam- shaping lenses L1 and L2 (Fig.~ \ref{fig1}(c)). The trapping chamber (GC) is a rectangular glass cuvette placed on a microscope glass slide affixed above the output pupil of MO1. The coated particles are taken on a glass cover slip that is attached on top of the sample chamber using sticky tape, so that the particles being on the inner surface fall down under gravity as the cover slip is perturbed mechanically. The focus of the objective is about 10 mm from the lower surface of GC. Imaging of trapped particles is performed on a CCD camera by a second 10X objective MO2 in a direction perpendicular to the trapping beam. The scaling of the images is performed by placing a known microscope calibration length standard (graduations at 10 $\mu m$ intervals) at the focal plane of MO2. Particles are typically trapped about 1 mm above or below the focal point (Fig.\~ref{fig1}(e)) as is usually the case in photophoretic trapping. We generally trap particle clusters which is clear from the Fig.~\ref{fig1}(f) that is the zoomed-in image of Fig. \ref{fig1}(e). The size of the trapped cluster is measured to be around 52 $\mu m$. We often observe a chain of trapped particles similar to that reported in \cite{zhang2012observation} - however, our focus in this paper is on single clusters. The motion of the trapped particle in the radial ($x$) direction is detected by a position sensitive detection system constructed around a balanced detection scheme \cite{li2010measurement} designed by imaging the scatter from the trapped cluster on the edge mirror EM using lens L3. The edge mirror splits the detection beam into two halves that are focused on detectors PD1 and PD2 (Thorlabs PDA100A-EC Si-photodiodes). The output of the detectors is fed into sum and difference amplifiers such that the normalized detection signal is given by $\dfrac{A-B}{A+B}$, where A and B are the outputs of the two detectors. Note that axial motion can be measured by rotating EM by $90\deg$ from the configuration shown in Fig.~\ref{fig1}(c). The final output is recorded in a computer using Labview. We record the Brownian motion of the trapped particle cluster at laser powers of 50, 100, and 200 mW before MO1. Note that as the laser power is increased, the axial trapping position is modified slightly towards higher values of $z_{0}$ as shown in Fig.~\ref{fig1}(a), so that the imaging as well as the detection systems have to be realigned to obtain maximum signal. 
\begin{figure}
\includegraphics[scale=0.25]{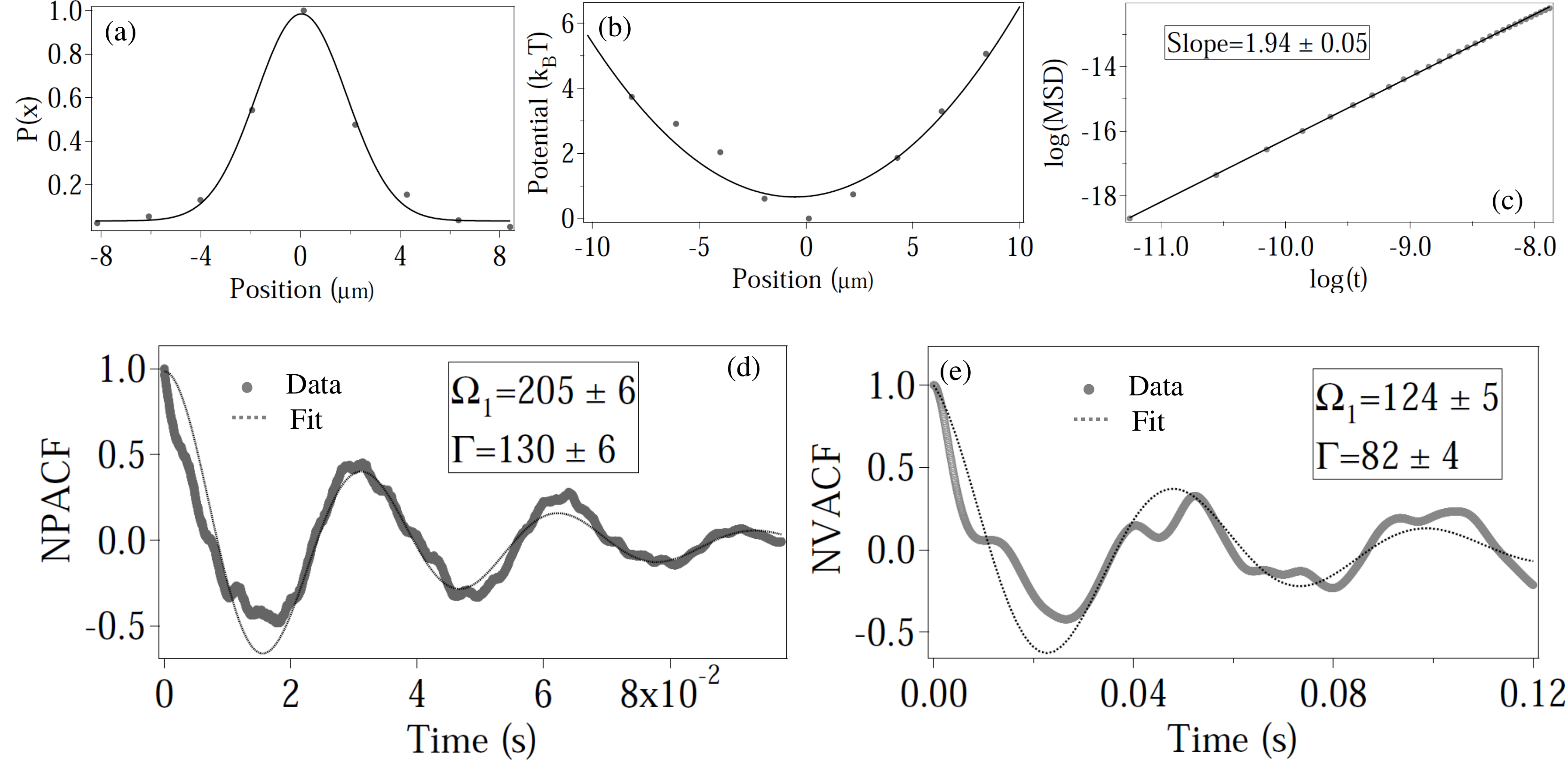}\caption{\label{fig3}(a) Nature of position probability distribution of $x$ component of Brownian motion fit to a Gaussian (solid line). (b) Nature of trapping potential determined from (a), fit to $y=cx^{2}$, where $c$ is a constant. (c) Log-log plot of MSD vs time. (d) $NPACF$ at a laser power of 200 mW. (e) $NVACF$ at a laser power of 50 mW.}
\end{figure}
\begin{figure}
\includegraphics[scale=0.28]{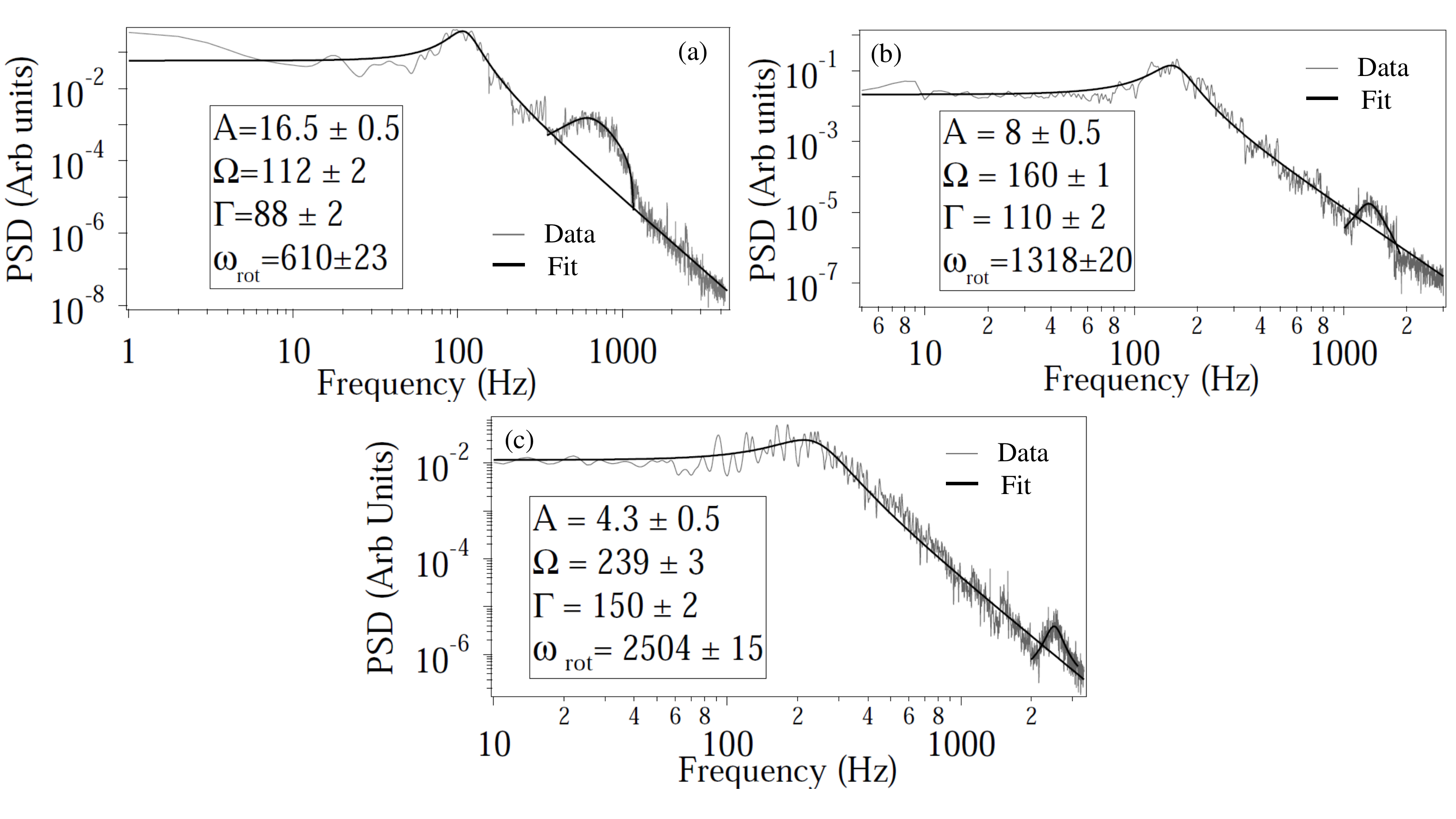}\caption{\label{fig4}PSD of trapped cluster at trapping powers of (a) 50 mW,
(b) 100 mW, (c) 200 mW. Data fit to Eq.~\ref{eq4}}
\end{figure}
A trapped Brownian particle under the influence of a linear restoring
force obeys a second order Langevin equation given by 
\begin{equation}
\ddot{x}+\Gamma\dot{x}+\Omega^{2}x=\Lambda\zeta(t),\label{eq1}
\end{equation}
where $\Gamma=\dfrac{\gamma}{m}$, $\gamma=6\pi\eta a$, $\eta$ being
the viscosity of air, $a$ the radius of a particle assuming it is
a sphere, and $m$ the mass. $\Omega$ is the natural frequency given
by $\Omega^{2}=\dfrac{k}{m}$, $k$ being the stiffness, $\Lambda=\sqrt{\dfrac{2k_{B}T\Gamma}{m}}$,
and $\zeta(t)$ is the delta-correlated stochastic noise characteristic
of Brownian motion. Note that $k$ in our case originates from $F_{\Delta\alpha}$,
with the latter being linearly proportional to the intensity $I$ of the trapping
laser \cite{Rohatschek1995,wurm2008} at a particular air pressure. This implies that $k$ too should vary linearly with intensity as we demonstrate later.
We proceed to determine the characteristics of the radial Brownian
motion in the time domain. The results are shown in Fig. \ref{fig3}.
Fig.~\ref{fig3}(a) shows the normalized position probability histogram $P(x)$ -  in the $x$ direction which fits very well to a Gaussian. Note however, that the signal also includes the contribution of the rotation of the center of mass of the particle cluster, so that the total extent of the position distribution is possibly amplified from the case of pure translational/rotational Brownian motion. However, we are more interested in the qualitative nature of the position probability distribution which is useful in understanding the basic properties of photophoretic trapping. We determine the effective trapping potential $E$ from $P(x)$ using the relation $E=-k_{B}T\ln P(x)$, where $k_{B}$ is the Boltzmann constant, and $T$ the temperature. The data fits well to a parabola of the form $y=cx^{2}$, where $c$ is a constant, which demonstrates that the effective potential near the trap center is parabolic as in the case of optical trapping so that the restoring force is clearly linear. Note again that the qualitative nature of the potential is of interest to us, so that the nature of the $x$-scaling is largely unimportant. The potential is clearly harmonic which is consistent with particle rotation as well (since rotation is also a manifestation of simple harmonic motion). It is worthwhile to point out here that the exact particle dynamics in the time domain can be determined by a fast camera which will be able to resolve the Brownian motion from the rotation of the particle - this, however, is presently not available with us. We next determine the mean squared
displacement ($MSD$) of the Brownian motion data and using the relation
$MSD=\dfrac{k_{B}T}{m}t^{2}$, $t$ being the time, we plot $\log MSD{\rm \;vs}\;\log t$
in Fig. \ref{fig3}(c). The value of the slope is $1.94\pm0.05$,
which signifies that we are indeed in the ballistic domain of Brownian
motion - where the expected slope is 2. Finally, we calculate the $NPACF$ and $NVACF$ of the data. Note that the instantaneous velocity of a trapped particle can be calculated only in the ballistic domain due to the $t^{2}$ dependence of the $MSD$. The $NPACF$ and $NVACF$ can be related to the parameters
$\Omega_{1}$ and $\Gamma$ by the following \cite{li2010measurement}
\begin{eqnarray}
NPACF=& \dfrac{\left\langle x(t)x(0)\right\rangle }{\left\langle x^{2}\right\rangle } =& \exp\left(-\dfrac{\Gamma t}{2}\right)\times \nonumber \\
&&\left(\cos\Omega_{1}t+\dfrac{\Gamma\sin\omega_{1}t}{2\omega_{1}}\right),\label{eq2}\\
NVACF=&\dfrac{\left\langle v(t)v(0)\right\rangle }{\left\langle v^{2}\right\rangle }=&\exp\left(-\dfrac{\Gamma t}{2}\right)\times\nonumber \\ &&\left(\cos\Omega_{1}t-\dfrac{\Gamma\sin\omega_{1}t}{2\omega_{1}}\right),\label{eq3}
\end{eqnarray}
where $\Omega_{1}=\sqrt{\Omega^{2}-\dfrac{\Gamma^{2}}{4}},$ $x$
and $v$ denote position and velocity, respectively. We determine
the $NPACF$ and $NVACF$ for all values of the trapping laser powers
used by directly computing the autocorrelations from the Brownian
motion data. For representation, we demonstrate the $NPACF$ at 200
mW (Figs. \ref{fig3}(d)), and the $NVACF$ at 50 mW (Figs. \ref{fig3}(e))
laser power. The calculated autocorrelations are each fit to the RHS
of Eqs. \ref{eq2} and \ref{eq3} for the position and velocity, respectively.
We determine the fit parameters $\Omega$ and $\Gamma$ (the errors
in the fit values are also shown at the $1\sigma$ level), from both
the autocorrelation functions at all laser powers and the values come
within 10\% of each other at each power. However, the fits are not particularly
good, and this may be due to the fact that the time series data is
a convolution of the Brownian motion and the rotation due to the body
force, which the fit functions fail to account for. This is why we
resort to the frequency domain to obtain clear signatures of the inertia
and the rotation, and use these values to calculate the mass of the
trapped particle cluster. We do compare the fit values of $\Omega$ and $\Gamma$ obtained by the time domain analysis to those obtained by the frequency domain analysis and observe agreement at the $2-3~\sigma$ level, as we show later.
\begin{table*}
\caption{\label{table1}Fitted values of the parameters $\Gamma$/m and $\Omega$
from PSD of the trapped particle at three different trapping laser
powers. Values of the coefficient of viscosity ($\eta$) are taken
from standard tables with the corresponding air-temperature values
considered for each laser power given in parenthesis. $m_{fit}$ and
$k$ are calculated in each case. }
\label{coeff_table} %
\begin{tabular}{|c|c|c|c|c|c|c|c|c|c|}
\hline 
Radius & Density & Mass & Laser & $\eta$ & Fitted & Fitted $\Omega$ & Mass from & Average & Calculated\tabularnewline
of particle ($a$) & $\rho$ & $\left(m_{D}\right)$ & power & (standard values)  & $\Gamma=\dfrac{\gamma}{m}$  & from PSD & fit$\left(m_{fit}\right)$  & $m_{fit}$ & k \tabularnewline
($\mu$m) & (kg/m\textsuperscript{3}) & (kg) & (mW) & (kg/m s) & from PSD &  & (kg)  &  (kg) & (N/m)\tabularnewline
\hline 
 &  &  & 50 & 1.96e-5 (325K) & 88(2)  & 112(2)  & 1.09(2)e-10 &  & 1.37(6)e-6\tabularnewline
26 & 1300 & 9.55e-11 & 100 & 2.18e-5 (375K) & 110(2)  & 160(1)  & 9.72(18)e-11 & 9.68(1.24)e-11 & 2.49(11)e-6\tabularnewline
 &  &  & 200 & 2.58e-5 (475K) & 150(2)  & 239(3)  & 8.42(12)e-11 &  & 4.81(17)e-6\tabularnewline
\hline 
\end{tabular}
\end{table*}
We now analyze the data in the frequency domain by determining the
PSD of the Brownian motion for the laser powers mentioned earlier.
The sampling frequency is 6.5 kHz, and we average 25 individual spectra
to generate each final spectrum. These are shown in Figs. \ref{fig4}(a)-(c).
The PSD is given by 
\begin{equation}
S(\omega)=\beta^{2}\dfrac{2k_{B}T}{k}\dfrac{\Omega^{2}\Gamma}{\left(\Omega^{2}-\omega^{2}\right)^{2}+\omega^{2}\Gamma^{2}}.\label{eq4}
\end{equation}
Here, $\beta^{2}$ is the conversion factor of the detector from voltage
to actual displacement. We do not determine $\beta^{2}$ in the present
case since it is not required in the measurements we report. We
fit Eq. \ref{eq4} to the data shown in Figs. \ref{fig4}(a)-(c). Now,
the data show clear peaks at the natural frequency and the rotation.
The fit function does not fit the rotation peak as expected, and we
separately fit Lorentzian functions to determine the centers of the rotation
peaks ($\omega_{rot})$. The fit parameters are shown in Table \ref{table1}.
It is clear that as the laser power is increased, the values of $\Omega,$ $\Gamma,$ and
$\omega_{rot}$ increase. However, while $\omega_{rot}$ increases
linearly within $3\sigma$, the rates of increase of $\Omega$ and
$\Gamma$ are different with $\Omega$ $\left(=\sqrt{\dfrac{k}{m}}\right)$
increasing by a factor of around $\sqrt{2}$ as the laser power is
doubled, while the increase in $\Gamma$ is proportional to the value
of $\eta$ in air as a function of temperature. The increase in $\Gamma$,
and accordingly the viscous damping as the laser power is increased
is evident in Fig \ref{fig4}(a)-(c) from the gradual broadening of
the resonance peak. From the fit values of $\Gamma$, we determine
$m$ by using $m=\dfrac{\gamma}{\Gamma}$, where the values of $\eta$
for calculating $\gamma$ are evaluated from standard tables \cite{engineering}
assuming particular temperature values at different laser powers.
We infer the temperature of the air in the following manner: we calculate
the mass $\left(m_{D}\right)$ of the particle cluster from the measured
diameter and the density which we estimate to be 1300 $kg/m^{3}$ from
the average thickness of the $PbS$ coating (100 $nm$) and
its density (7600 $kg/m^{3}$) compared to the average diameter and
density of the $SiO_{2}$ particles. For the lowest laser power of
50 mW measured near MO1, we consult the viscosity tables and select that value of viscosity
using which the value of mass $\left(m_{fit}\right)$ from the fit
value of $\Gamma$ is reasonably close to $m_{D}$. This occurs at
around 325K, which implies that the laser heating of the particle
has led to an increase of the air temperature in its vicinity by around
25K from room temperature, which is not very unreasonable. With
this value of $\eta$, $m_{fit}$= 1.09(2)e-11 kg - the $1\sigma$
error in parenthesis being due to the error in the fit for $\Gamma$.
This is within 15\% of $m_{D}$. For consistency check, we use the
two other laser power values (100 and 200 mW), and assuming a linear
dependence of air temperature increase with laser power, we find (Table
1) that the values of $m_{fit}$ are within 10\% of $m_{D}$. Note
here, that even if we assume a constant value of $\eta$ for the different
laser power values, the value of $m_{fit}$ differs only by 33\% from
$m_{D}$, which implies that our error estimates are not unreasonable. The average value of $m_{fit}$ is $9.68\pm 1.24~e-11$ which is indeed very close to $m_D$, albeit with a $1\sigma$ error of around 15\%.  We now calculate the stiffness $k$ at the different laser powers and observe from Table 1 that as expected, it increases linearly within the $1\sigma$ error values. In addition, the amplitude of the power spectra, $A = \beta^{2}\dfrac{2k_{B}T}{k}$, decreases linearly as $k$ increases, which again acts as a consistency check to the data. Finally, we observe that the values of $\Gamma$ and $\Omega$ obtained from the $NPACF$ and $NVACF$ shown in Fig. \ref{fig3}(d) and (e) agree with that from the PSD at the same laser powers at the $2-3\sigma$ level.

In conclusion, we develop a very simple optical trapping set-up for
confining absorbing particles in air with a single loosely focused Gaussian laser beam. We characterize the radial component of the photophoretic body force
$F_{\Delta\alpha T}$ and show its equivalence to the optical intensity
gradient force commonly used in optical tweezers. For this, we detect
the radial ($x$) Brownian motion and analyse it both in the time and frequency
domain. The latter seems the most reliable
technique to study the motion of trapped particles, as it clearly
separates out the inertial resonance from the rotation induced by $F_{\Delta\alpha T}$.
We find out the changes in both frequencies due to increase in laser
power, and are also able to extract an estimate of the particle mass
with around 15\% accuracy by fitting the power spectral density to
the analytical expression derived from the Langevin equation - a procedure we intend to improve in the future by employing Bayesian statistics. Ours
is the first direct characterization of particle motion induced by
photophoretic forces using a very simple experimental set-up, and
may set the path for more precise experiments that could help develop
crucial understanding about photophoretic forces that have deep connotations
in diverse natural phenomenon ranging from planet formation \cite{planetform2013} to stratification in the atmosphere \cite{wurm2008}. 

This work was supported by the Indian Institute of Science Education
and Research, Kolkata, an autonomous research and teaching institute
funded by the Ministry of Human Resource Development, Govt. of India.

{\bf Complete References:}
\begin{enumerate}
\item O. Jovanovic, ``Light induced motion of particles suspended in gas", J.  Quant. Spectroscop. and Rad. Transf. {\bf 110}, 889--901 (2009).
\item A. S. Desyatnikov, V. G. Shvedov, A. V. Rode, W. Krolikowski, and Y. S. Kivshar ``Photophoretic manipulation of absorbing aerosol particles with vortex beams:
theory versus experiment",  Opt. Express \textbf{17},  8201--8211 (2009).
\item M. Braun and F. Cichos, ``Optically controlled thermophoretic trapping of single nano-objects",  ACS nano \textbf{7}, 11200--11208 (2013).
\item  L. Jauffred, S. M.-R. Taheri, R. Schmitt, H. Linke, and
L. B. Oddershede, ``Optical trapping of gold nanoparticles in air", Nano Lett.  \textbf{15}, 4713--4719 (2015).
\item V. G. Shvedov, A. V. Rode, Y. V. Izdebskaya, A. S.
Desyatnikov, W. Krolikowski, and Y. S. Kivshar, ``Giant optical manipulation", Phys. Rev.  Lett. \textbf{105}, 118103 (2010).
\item V. G. Shvedov, A. S. Desyatnikov, A. V. Rode, W. Krolikowski, and Y. S. Kivshar, ``Optical guiding of absorbing nanoclusters in air",  Opt.  Express \textbf{17}, 5743--5757 (2009).
\item Z. Zhang, D. Cannan, J. Liu, P. Zhang, D. N.
Christodoulides, and Z. Chen, ``Observation of trapping and transporting air-borne absorbing particles with a single optical beam", Opt. Express \textbf{20},  16212--16217 (2012).
\item J. Lin and Y. qing Li, ``Optical trapping and rotation of
airborne absorbing particles with a single focused laser
beam", Appl. Phys.  Lett. \textbf{104}, 101909 (2014).
\item H. Rohatschek, ``Semi-empirical model of photophoretic
forces for the entire range of pressures", J. Aerosc. Sc. \textbf{26}, 717--734
  (1995).
\item G. Wurm and O. Krauss, ``Experiments on negative photophoresis and application to the atmosphere", Atmosph. Env. \textbf{42},  2682--2690 (2008).
\item T. Li, S. Kheifets, D. Medellin, and M. G. Raizen, ``Measurement of the instantaneous velocity of a brownian
particle", Science \textbf{328},  1673--1675 (2010).
\item http://www.engineeringtoolbox.com/dry-air-properties-d\_973.html.
\item J. Teiser and S. E. Dodson-Robinson, ``Photophoresis
boosts giant planet formation", Astronom. and Astro-
phys. \textbf{555}, A98 (2013).

\end{enumerate}

\end{document}